# Formal requirement and architecture specifications of a multi-agent robotic system

Nadeem Akhtar, Yann Le Guyadec, and Flavio Oquendo

**Abstract**— One of the most challenging tasks in specification engineering for a multi-agent robotic system is to formally specify and architect the system, especially as a multi-agent robotic system is concurrent having concurrent processing, and often having dynamic environment. The formal requirement and architecture specifications along with step-wise refinement from abstract to concrete concepts play major role in formalizing the system. This paper proposes the formal requirement and architecture specifications aspects of an approach that supports analysis with respect to functional as well as non-functional properties by step-wise refinement from abstract to concrete specifications and formal architecture definition. These formal specifications have been exemplified by a case study. As formal specification techniques are getting more mature, our capability to build a correct complex multi-agent robotic system also grows quickly.

—————————— ◆ ——————————

## 1 INTRODUCTION

ONE of the most challenging tasks in specification engineering of a multi-agent robotic system is to formally specify and architect the system, especially as it is concurrent having concurrent processes, and often having dynamic environment. An approach has been proposed that supports formal analysis with respect to functional as well as non-functional properties; that supports step-wise refinement from abstract to concrete specifications and full code generation; and that formalizes the static as well as the dynamic architecture of these systems.

An agent is a computer system situated in some environment, capable of autonomous actions in this environment in order to meet its design objectives [16]. Multiple agents are necessary to solve a problem, especially when the problem involves distributed data, knowledge, or control. A multi-agent system is a collection of several interacting agents in which each agent has incomplete infor-mation or capabilities for solving the problem [5]. These are complex systems and their specifications involve many levels of abstractions.

We have proposed a formal approach having four phases of requirement specifications, verifica-tion specifications, architecture specifications, and system implementation as shown in fig. 1. This approach identifies and formally specifies each component and sub-component of the system, and identifies the formal requirement specifi-cations; verification specifications; architecture specifications; implementation; satisfaction and refinement relations between different phases. This ap-proach has been exemplified by a case study of a multi-agent robotic system. Complete approach with its functional and technical details can be found in [1].

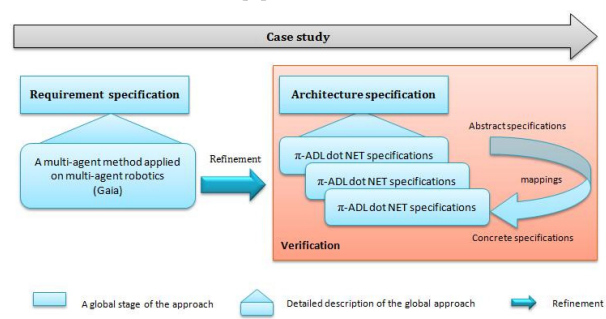

Fig. 1. The proposed approach having two major phases of requirement and architecture specification.

This paper focuses on the formal requirement and architecture specifications. Our contributions are; (1) a combination of regular expression, first-order predicate logic to define the formal requirement specifications; (2) π-calculus based π-ADL dot NET platform to define the dynamic architecture specifications; (3) a multi-agent robotic system case study to exemplify these formal specifications.

Objective: Our major objective is the formal specification, analysis with respect to functional as well as non-functional properties by step-wise refinement from abstract to concrete specifications and then formal architecture specifications. Along with the development of a robotic multi-agent system that's static as well as dynamic formal architecture can be defined and is safe. By safe the focus is on the correctness properties of safety and liveness.

Section 2 presents the background studies, Section 3

————————————————

- *Nadeem Akhtar is an assistant professor at the Department of Computer Science & IT, The Islamia University of Bahawalpur, PAKISTAN, E-mail:akhtarnadeem@hotmail.com.*
- *Yann Le-Guyadec is an associate professor at the Laboratory VALORIA of Computer Science, University of South Brittany (UBS), Vannes, FRANCE, E-mail: yann.le-guyadec@univ-ubs.fr.*
- *Flavio Oquendo is a Professor at the Laboratory VALORIA of Computer Science, University of South Brittany (UBS), Vannes, FRANCE, E-mail: flavio.oquendo@univ-ubs.fr*

the case study, and Section 4 lessons learned and conclusion.

## 2 BACKGROUND STUDIES

### 2.1 Gaia multi-agent method

Among the existing multi-agent methods we have considered Gaia [17] as the most suitable one for specifying the requirements as it recognizes the organizational structure as a primary dimension for the development of an agent system. This organizational structure provides organizational abstractions, which are needed to meet both functional and non-functional requirements. Gaia considers a multi-agent system as a computational organization consisting of interacting roles, and it deals with both the macro-level (social) and the micro-level (agent internal) aspects of a multi-agent system.

The analysis phase in Gaia specifies the requirements in terms of functions, activities, and identi-fies the loosely coupled sub-organizations which compose the whole system. It involves considering real-world organization, the need to enforce organizational rules and then, for each of these sub-organizations it produces the following basic abstract models:

(1) Environmental model captures the characteristics of the multi-agent system operational environment.

(2) A preliminary role model captures the key task-oriented activities to be played in a multi-agent system. The result is a prototypical role model, a list of the key roles that occur in the system, each with a description that is not elaborated.

(3) A preliminary interaction model captures basic inter-dependencies between roles. The result is an interaction model, which captures the recurring patterns of role interactions.

(4) The preliminary interaction model is used as a base to elaborate the roles. The result is a fully elaborated roles model, which documents the key roles of the system, their permissions and responsibilities. Responsibilities attributes determine the expected behavior and key attributes of liveness and safety associated with a role. Liveness property is specified via liveness expression which defines the potential execution trajectories through activities and protocols associated with the role. An activity corresponds to a unit of action that does not involve interaction with any other agent, and protocol requires interaction with other agents.

(5) A set of organizational rules, expressing global constraints or directives that underlie the multi-agent system functioning. The role and interaction models are completed based on the adopted organizational structure. The analysis models are input to architectural design phase which defines the organizational structure of the system. After the architectural design phase, the detailed design involves identifying: An agent model consisting of a set of agent classes in a multi-agent system, implementing the identified roles; and a services model, expressing services and interaction protocols to be provided within these agent classes.

### 2.2 π-ADL

Formal requirements and architecture specifications achieve precision; the unambiguous specifications ensure that correctness, completeness, and complex system properties are preserved.

π-ADL [11] provides the core structure and behavior constructs for describing static as well as dynamic software architectures. It is a formal specification language designed to be executable and to support automated analysis and refinement of dynamic architectures. It has as formal foundation the higher-order typed π-calculus [6][7][8][13], and it takes its roots in work concerning the use of π-calculus as a semantic foundation for architecture description languages [3][4]. The design of π-ADL are based on [9][14][15] and follows the language design principles found in [10]. π-ADL supports description of software architectures from a runtime perspective, an architecture is described in terms of components, connectors, and their composition. Fig.2 depicts its main constituents.

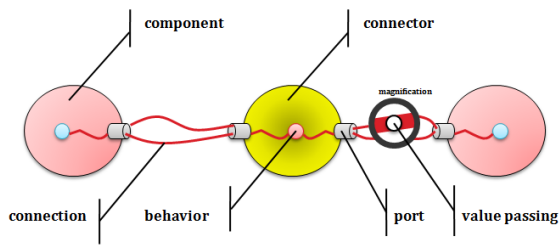

Fig 2. Architectural concepts in π-ADL [11]

Components are described in terms of external ports and an internal behavior. Their architectural role is to specify computational elements of a software system. The focus is on computation to deliver system functionalities. Ports are described in terms of connections between a component and its environment. Their architectural role is to put together connections providing an interface between the component and its environment. Protocols may be enforced by ports and among ports. Connections are basic interaction points. Their architectural role is to provide communication channels between two architectural elements. A component can send or receive values via connections. They can be declared as output connections (values can only be sent), input connections (values can only be received), or input-output connections (values can be sent or received). Connectors are special-purpose components. They are described as components in terms of external ports and an internal behavior. However, their architectural role is to connect together components. They specify interactions among components. Therefore, components provide the locus of computation, while connectors manage interaction among components. A component cannot be directly connected to another component. In order to have actual communication between two components, there must be a connector between them. Both components and connectors comprise ports and behaviour. In order to attach a port of a component to a port of a connector, at least a connection of the former port must be attached with a connection of the later port. A connection provided by a port of a component is attached to a connection provided by a port of a connector by unification or value passing. Thereby, attached connections can transport values (that can be data, connections, or even architectural elements).

π-ADL dot NET [12] is the dot NET extension of π-ADL based on Microsoft dot NET platform. It provides an executable model of system specifications consisting of abstractions and behaviors, and leads to a formal architecture comprising of components and connectors that can change dynamically dusring execution.

(a) It has as formal foundation the higher-order typed π-calculus [6][7][8][13] which is a well-formed higher-order calculus for defining communicating and mobile architectural elements; (b) It focuses on formal description of software architecture from the run-time viewpoint: the run-time structure, the run-time behavior, and how it may evolve over time; (c) It is executable; (d) It supports multiple concrete syntaxes: textual and graphical notations; and (e) It supports automated verification of properties by model checking and theorem proving.

## 3 CASE STUDY: MULTI-AGENT ROBOTIC TRANSPORT SYSTEM

Our multi-agent robotic system is composed of transporting agents. The mission is to transport stock from one storehouse to another. They move in their environment which is static i.e. the topology of the system does not evolve at run time [2][1].

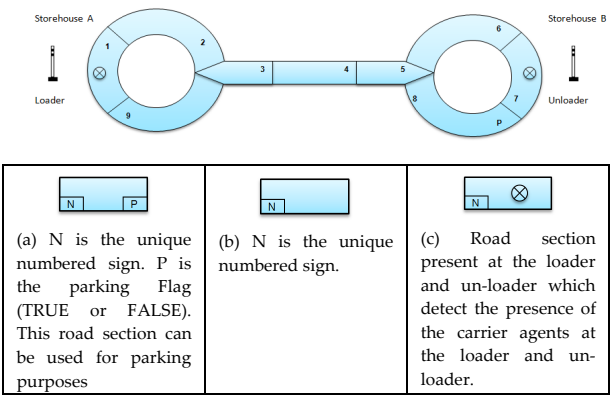

(a) N is the unique numbered sign. P is the parking Flag (TRUE or FALSE). This road section can be used for parking purposes

(b) N is the unique numbered sign.

(c) Road section present at the loader and un-loader which detect the presence of the carrier agents at the loader and un-loader.

Fig. 3. Case study - Environment alongwith the loader and unloader agents.

Our system consists of three types of agents;

(1) Carrier agent transports stock from one storehouse to another one, it can be loaded or un-loaded; can move both forward and backward directions; and can detect collision. Each road section is marked by a sign number readable by the carrier agent.

(2) Loader / Un-loader agent receives/delivers stock from the storehouse; ensures that the carrier waiting to be loaded is loaded and the carrier waiting to be un-loaded is unloaded.

(3) Store-manager agent manages the stock count in the storehouse and transports the stock between the storehouse and loader/un-loader.

*Environment:* There is a path between storehouse-A and storehouse-B which is composed of a sequence of interconnected road sections of fixed length as shown by fig.2. Each road section has a numbered sign, which is readable by carrier agents. Each road section has a unique numbered sign. There are three types of road sections depending upon the road topology. The road is single lane and there is a roundabout at storehouse-A and storehouse-B [2].

*Scenario:* In the case study we have used a particular road topology consisting of nine road partitions as shown in fig.3. The main task of the carrier is to transport the stock from storehouse A to storehouse B until the storehouse A is empty. Loader at the storehouse A loads the carrier with stock and the Un-loader at the storehouse B unloads the carrier. The store-manager keeps a count of stock in each storehouse.

### 3.1 Requirement specifications based on Gaia

The role of an agent defines what it is expected to do in the organization, both in concert with other agents and in respect to the organization itself. Organizational role model precisely describes each role that constitutes the computational organization. Here we present the Move_full role of our system.

TABLE 1
MOVE_FULL ROLE OF GAIA ROLE MODEL

| Role Schema: **Move_full** |
|---|
| **Description:** |
| Role of a loaded carrier moving from storehouse A to storehouse B. |
| **Protocols and Activities:** |
| readSign, movetoNext, collisionSensorTrue, carrierWait, readUnloadSign, waitforUnloading, unloadCarrier |
| **Permissions:** |
| reads: *sign_number (external)* |
| *collision_sensor (internal)* |
| changes: *position (internal)* |
| *next_position (external)* /// (TRUE or FALSE) Checks if next position is available |
| **Responsibilities:** |
| *Liveness:* |
| **Move_full** = |
| ***Move***.(readUnloadSign.waitforUnloading.unloadCarrier) |
| ***Move*** = (readSign.movetoNext)+ |
| \| (collisionSensorTrue.***Wait***).(readSign.movetoNext)+ |
| ***Wait*** = carrierWait+ |
| *Safety:* |
| is_Full(c) ∧ can_movetoNext(sn) |
| where c is for carrier and sn for the sign number |

In the above table activities (underlined) are readSign, movetoNext, collisionSensorTrue, carrierWait and readUnloadSign. And there are two protocols waitforUnloading and unloadCarrier. The activities are actions of an agent that do not involve interaction with any other agent, whereas protocols are actions that require interaction with other agents. When a loaded carrier reaches the road partition in front of the un-loader, it stops there and waits until it is unloaded. When we consider the liveness property, it shows all the activities and protocols that make up the role. The carrier has two choices; First choice it reads the sign and then moves to the next road partition; Second choice in case of collision with another carrier it sets its collision sensor to true and then waits. At the end, the carrier reads the unload sign, waits to be unloaded, so now it's no longer a loaded carrier, and is therefore not part of the Move_full role, instead it is part of the Move_empty role.

There are dependencies and relationships between the various roles in a multi-agent organization which are protocol definitions, one for each type of inter-role interaction. Table.2 shows the protocol definitions related to Move_full role.

TABLE 2
PROTOCOL DEFINITIONS RELATED TO MOVE_FULL ROLE

| waitForUnloading | | |
|---|---|---|
| Move_full | Unload | *sign_number* |
| The full carrier agent waits for the un-loader agent | | *position* |

| unloadCarrier | | |
|---|---|---|
| Move_full | Unload | *sign_number* |
| The full carrier agent is unloaded by the un-loader agent | | *position* |

### 3.2 Architecture specifications based on π-ADL dot NET

π-ADL dot NET architecture encompass both behavioral and architecture-centric constructs, specify static along with dynamic architectural elements, and achieve Turing completeness and high architecture expressiveness with a simple formal notation. Here we present the architecture definition of the case study presented in section 3. Initially storehouse-A is full and storehouse-B is empty. This π-ADL dot NET based architecture elaborates static as well as dynamic aspects of the architecture.

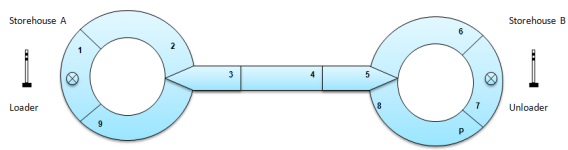

Fig. 4. Case study - Environment.

The π-ADL architecture model of this system is shown in fig. 5. There is one behavior ROUTE and four abstractions MOVE_FULL, MOVE_EMPTY, STOREHOUSE_A and STOREHOUSE_B. There is a connection of type connection (i.e. connection that lets another connection pass through it) between MOVE_FULL and STOREHOUSE_A, similarly another connection of type connection between MOVE_EMPTY and STOREHOUSE_B. These connections of type connection specify the dynamic aspects of the architecture.

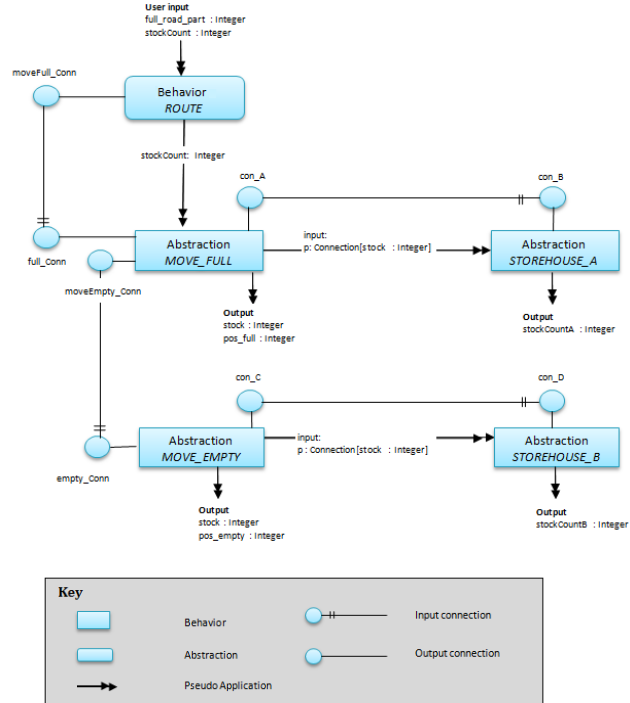

Fig. 5. π-ADL dot NET model of our system.

This π-ADL architecture model as shown in fig. 5 is then defined by the π-ADL dot NET code which can be compiled and run, some part of this code is given below. ROUTE is the main behavior having one connection named moveFull_Conn, through which it sends the road partition no. of full carrier agent to the MOVE_FULL abstraction. The no. of stock present in storehouse-A (stockCount) is sent as an argument to MOVE_FULL through connection renaming.

```
1   /// ROUTE is the main bahaviour i.e. from where the execution
2   /// starts
3   ROUTE names behaviour
4   {
5   moveFull_Conn: Connection[Integer];
6   full_road_part: Integer;
7   stockCount: Integer;
8   /// Gets the initial location of the full carrier
9   via in receive full_road_part;
10  /// Gets the number of stock to transfer from STOREHOUSE_A to
11  /// STOREHOUSE_B
12  via in receive stockCount;
13  compose{
14  via MOVE_FULL send stockCount where
15       {moveFull_Conn renames full_Conn};  /// sends stockCount
16  and
17  /// sends full carrier initial position i.e. road partition
18  via  moveFull_Conn  send  full_road_part;       } /// end compose
19  } /// end ROUTE
```

MOVE_FULL refers to the role of a loaded carrier agent. It has a connection named moveEmp-ty_Conn, and through this connection it sends the road partition no. of loaded carrier to MOVE_EMPTY. There is another connection con_A which allows another connection to pass through it. Through this connection a connection is sent with the current number of stock from MOVE_FULL to STOREHOUSE_A. So we have an architecture component (i.e. connection) that can pass other architecture components through it during execution, thus changing the architecture at runtime, as a result this con_A shows one of the dynamic aspects of our architecture.

```
20   /// Here its the MOVE_FULL role as an abstractions (single
21   /// carrier agent role)
22   value MOVE_FULL is abstraction(x: Integer)
23   {
24   pos_full:Integer;
25   empty_road_part: Integer;
26   stock: Integer;
27   a: Integer;
28   moveEmpty_Conn: Connection[Integer];
29   full_Conn: Connection[Integer];
30   /// con_A is a dynamic connection between the MOVE_FULL carrier
31   /// role and storehouse-A
32   /// Connection passed through a connection (dynamicity of the
33   ///system)
34   con_A: Connection[Connection[Integer]];
35   p: Connection[Integer];
36   stock = x;
37   via full_Conn receive x;
38   pos_full = x;
39   /// If full carrier is out of the range of full road partitions
40   if(pos_full<1 || pos_full>7) do
41       { via out send "Invalid position for Full carrier";
42   done;         }
43   /// If stock is zero than the system stops as carrier has no stock to
44   /// transfer
45   if(stock==0) do
46       { via out send "No stock to transfer";
47   done;         }
48   while(pos_full>=1 && pos_full<=6) do{
49   ///   pseudo-code here (due to space constraints)
50   ///   if(pos_full==1) do
51   ///   road position=1 is the loader location, move to next
52   ///   position and print the new position
53   ///   if (pos_full>1 && pos_full<=6) do
54   ///   road position 1 to 6 are the central route positions as shown
55   ///   in fig. 3
56   ///        move to the next position and print the new position
57   } /// end while
58   if(pos_full==7) do{
59   /// road position=7 is the Unloader location as shown in fig. 3
60   empty_road_part=8;
61   compose {
62   via STOREHOUSE_A send p where {con_A renames con_B};
63     and
64   /// connection A passes the connection p (i.e. fig. 4)
65   via con_A send p;
66   /// connection p sends the stock count as argument (i.e. fig. 4)
67   via p send stock;
68   compose {
69   via MOVE_EMPTY send stock where
70                 {moveEmpty_Conn renames empty_Conn};
71     and
72   via moveEmpty_Conn send empty_road_part;
73       } /// end compose
74   } /// end compose          } /// end if
75   } /// end MOVE_FULL
```

STOREHOUSE_A abstraction counts the number of stock and gives the state of the storehouse-A.

```
76   /// STOREHOUSE_A which is full initially and empty at the end
77   value STOREHOUSE_A is abstraction(input: Connection[Integer])
78   {
79   stockCountA:Integer;
80   con_B: Connection[Connection[Integer]];
81   via out send "Empty carrier at the LOADER";
82   via con_B receive input;
83   via input receive stockCountA;
84   via out send "No of stock at STOREHOUSE_A = ";
85   via out send stockCountA;
86   if(stockCountA > 0) do{
87     via out send "decrementStock_at_STOREHOUSE_A";
88     /// stock at the Storehouse-A is decremented by 1
89   stockCountA = stockCountA-1;      }
90   /// stock at the Storehouse-A can not be decremented as it is
91   ///already zero
92   if(stockCountA == 0) do
93       via out send "STOREHOUSE_A_Empty";
94   } /// end STOREHOUSE_A
```

## 4 LESSONS LEARNED AND CONCLUSION

The requirement and architecture aspects of a multi-agent robotic system are defined and formalized. The requirement specifications define the behavior alongwith the correctness properties of liveness and safety, and the π-ADL dot NET defines the formal static as well as the dynamic aspects of the architecture. This requirement specifications method has a concrete syntax to express properties, is suitable to model behaviors and is applicable to a wide range of multi-agent systems. These requirement specifications are refined into π-ADL dot NET based architecture specifications. The proposed formal approach has key aspects of: requirement specifications constituting of organizational abstractions, organizational rules, role model specifications, protocol definitions followed by the architecture specifications constituting of constructs for specifying static as well as dynamic architecture. This approach is a step towards the development of a method, centered on organizational abstractions and formal dynamic architecture for requirement analysis and architecture of a multi-agent robotic system.

**Nadeem Akhtar** is an Assistant Professor at the Department of Computer Science & IT, The Islamia University of Bahawalpur (IUB), Pakistan. Before joining IUB in 2011, he was a Ph.D fellow at the research Lab. VALORIA of Computer Science, University of South Brittany (UBS), France. He has been awarded a highly honoured Ph.D in Computer Science from Lab. VALORIA, University of South Brittany (UBS) France in 2010. He has an MS with specialization in Information system architecture from IUP, University of South Brittany, Vannes, France awarded in 2005. In 2004 he was awarded the French govt. "Study in France 2004" scholarship for post-graduate studies in France. In 2007 he was awarded a French govt. and Pakistan govt. HEC scholarship for PhD studies at lab. VALORIA-UBS, France. His research areas are formal architecture, formal specification, and service-oriented architecture for robotics.

**Yann Le-Guyadec** is an Associate Professor of Computer Science and Software Engineering at the University of South Brittany, part of the European University of Brittany, France. His research interests are in parallel processes, formal verification, and robotics.

**Flavio Oquendo** is a Full Professor of Computer Science and Software Engineering at the University of South Brittany, part of the European University of Brittany, France, where he leads the ARCH-LOG research team on Software Architecture at the VALORIA Research Lab, addressing languages, processes, and tools for designing, constructing and evolving software-intensive systems based on service-oriented architectures and component-based technlogies, targeting applications on different ICT domains and platforms. is a member of the IEEE and the IEEE Computer Society. He received the B.Sc. Eng. degree from the ITA Engineering School, Sao Paulo, and the M.Sc., Ph.D., and H.D.R. (Research Direction Habilitation) degrees from the University of Grenoble, in Computer Science. Before joining the University South Brittany in 2005, Prof. Oquendo held appointments in Computer Science & Software Engineering as Full Professor at the University of Savoie at Annecy (9 years) and Associate/Assistant Professor at the University of Grenoble (5 years). Previously, he was a member of the R&D Staff at the G.I.E. Emeraude [R&D group on advanced software engineering created by BULL, Thomson-Syseca (now THALES IS), and Eurosoft (now part of AT&T Europe) in Paris (5 years)].